\documentclass[conference]{IEEEtran}
\makeatletter
\def\ps@headings{%
\def\@oddhead{\mbox{}\scriptsize\rightmark \hfil \thepage}%
\def\@evenhead{\scriptsize\thepage \hfil \leftmark\mbox{}}%
\def\@oddfoot{}%
\def\@evenfoot{}}
\makeatother
\usepackage[latin1]{inputenc}
\usepackage[english]{babel}
\usepackage{dsfont}
\usepackage{amsfonts}
\usepackage{amssymb}
\usepackage{amsmath}
\usepackage{mathtools}
\usepackage{mathrsfs}
\usepackage{xcolor}
\usepackage{graphicx}
\usepackage{float}
\usepackage{stmaryrd}
\usepackage[colorinlistoftodos]{todonotes}
\usepackage[ruled,vlined]{algorithm2e}
\usepackage{verbatim}
\usepackage{color}
\usepackage{lipsum}
\usepackage{mathtools}
\usepackage{paralist}   

\usepackage{geometry}
\usepackage{tikz}
\usetikzlibrary{shapes,arrows}
\usetikzlibrary{calc}

\usepackage{url}

\pdfoutput=1

\newtheorem{theo}{Theorem}

\newtheorem{lem}{Lemma}
\newtheorem{cor}{Corollary}
\newtheorem{rem}{Remark}
\newtheorem{defi}{Definition}

\makeatletter
\renewcommand*\env@matrix[1][*\c@MaxMatrixCols c]{%
  \hskip -\arraycolsep
  \let\@ifnextchar\new@ifnextchar
  \array{#1}}
\makeatother

\IEEEoverridecommandlockouts

\title{Secure Strong Coordination}

\author{ 
  \IEEEauthorblockN{Giulia~Cervia, Germ{\'a}n~Bassi, and Mikael~Skoglund}
  
  \IEEEauthorblockA{School of Electrical Engineering and Computer Science,
                    KTH Royal Institute of Technology,\\
                    Stockholm, Sweden,
                    \{cervia, germanb, skoglund\}@kth.se}
  \thanks{This work was supported by the Knut and Alice Wallenberg Foundation and the Swedish Foundation for Strategic Research.}
}

\begin{document}

\maketitle

\begin{abstract} 
We consider a network of two nodes separated by a noisy channel, in which the source and its reconstruction have to be strongly coordinated, while simultaneously satisfying the strong secrecy condition with respect to an outside observer of the noisy channel. 
In the case of non-causal encoding and decoding, we propose a joint source-channel coding scheme for the secure strong coordination region.
Furthermore, we provide a complete characterization of the secure strong coordination region when the decoder has to reliably reconstruct the source sequence and the legitimate channel is more capable than the channel of the eavesdropper.
\end{abstract}

\section{Introduction}

Information-theoretic security and privacy has been an active field of
research for decades, with increased activity over the past 10--15
years~\cite{bloch2011physical}. More recently, information-theoretic
notions and tools have been applied also to related problems in
decision and control~\cite{Takashi-cloud} and learning \cite{Bae}.
This motivates research on new results in the information-theoretic
core that are valid beyond traditional transmission and storage of
information. Coordination was introduced in~\cite{cuff2009thesis} as
one such generalization of traditional information exchange problems,
with application in scenarios where action needs to be taken to align
statistical information over a network. However, so far there are
relatively few results on coordination with security constraints, with
the exception of~\cite{satpathy2014secure, satpathy2016secure} that
looked at secure coordination with noiseless links.

In particular, \cite{satpathy2014secure} considers the weaker metric of \emph{empirical coordination}, which requires the joint histogram of the actions in the network to be close with high probability to a desired distribution.
While empirical coordination is only interested in controlling the joint histogram, \emph{strong coordination} deals instead with the joint probability distribution of the actions. Whenever the average behavior over time is the concern, looking at the empirical joint distribution is enough. On the other hand, if an adversary is involved, a sequence of strongly coordinated actions appears truly random to an outside observer. In~\cite{satpathy2016secure}, the authors consider strong coordination for the case of noiseless links and an eavesdropper that receives the same message as the legitimate receiver.
However, since real-life communication is noisy, the assumption of an error-free link between the agents as in~\cite{satpathy2014secure,satpathy2016secure} should be revisited.

\begin{figure}[t]
 \centering
 \includegraphics[scale=0.2]{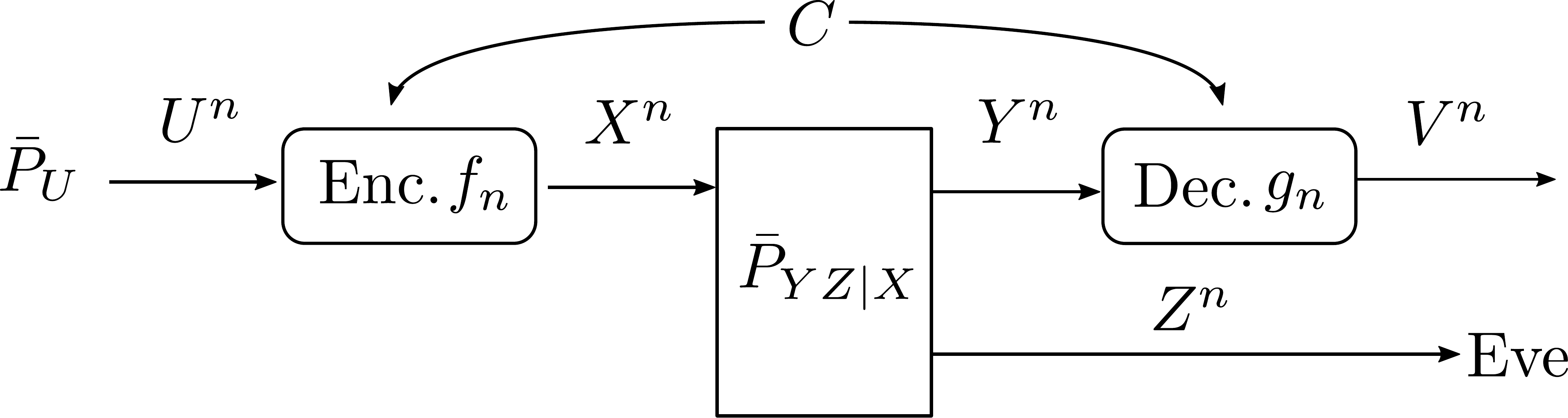}
\caption{System model. Eve observes the channel realization $Z^n$ but is oblivious of the common randomness $C$.}
\label{fig:wtap_model}
\end{figure}

In this paper, we study the problem of \emph{secure strong coordination} through a noisy channel and in the presence of an eavesdropper, as depicted in Fig.~\ref{fig:wtap_model}.
More precisely, we study a three-node network model comprised of an information source and a noisy channel, in which two agents, an encoder and a decoder, have access to a common source of randomness. Moreover, we consider an eavesdropper that observes an output of the noisy channel (possibly different from the decoder's) but has no knowledge of the common randomness.
This scenario presents two different goals: the encoder needs to convey a message to the decoder to strongly coordinate the reconstructed version of the source, while simultaneously ensuring that the eavesdropper is completely oblivious of the source and its reconstruction. 

We derive an inner bound for the secure strong coordination region by developing a joint source-channel scheme in which we introduce an auxiliary codebook that allows us to satisfy both requests. 
Although the region is still unknown in the general case, we are able to characterize the region when the decoder reliably reconstructs the source and the legitimate channel is more capable than the eavesdropper's channel.

The rest of the document is organized as follows. 
Section~\ref{sec: prel} introduces the notation and some preliminary results, while Section~\ref{sec: model} describes the model under investigation, states the main results, and compares the obtained original results with previous works.
Finally, the proofs of the main results are found in Section~\ref{sec: gen proof}.

\section{Preliminaries}
\label{sec: prel}

We define the integer interval $\llbracket a,b \rrbracket$ as the set of integers between $a$ and $b$.
Given a random vector $X^{n}\coloneqq$ $(X_1, \ldots, X_{n})$, we denote $X^{i}$ as the first $i$ components of $X^{n}$, and $X_{\sim i}$ as the vector $(X_j)_{j \neq i}$, $j\in \llbracket 1,n \rrbracket $, i.e., $X^{n}$ without the component $X_i$. 
We use $\mathbb V (\cdot , \cdot)$ and $\mathbb D (\cdot \Arrowvert \cdot)$ to denote the  total variation distance (or variational distance) and the Kullback--Leibler divergence between two distributions, respectively.
The notation $f(\varepsilon)$ denotes a function which tends to zero as $\varepsilon$ does, and the notation $\delta(n)$ denotes a function which tends to zero exponentially as $n$ goes to infinity.

We now state without proof some results that we need later.

\vspace{1mm}
\begin{lem}[Properties of the total variation distance]\label{tv prop}
\begin{enumerate}[(i)]
 \item \label{cuff16} $\mathbb V (P_{A}, \hat P_{A}) \leq \mathbb V (P_{AB}, \hat P_{AB})$, see~\cite[Lemma 16]{cuff2009thesis};
 \item \label{cuff17} $\mathbb V (P_A, \hat P_A)= \mathbb V (P_AP_{B|A}, \hat P_A P_{B|A})$, see~\cite[Lemma 17]{cuff2009thesis};%
 \item \label{lem4} if $ \mathbb V (P_{A} P_{B|A}, P'_{A} P'_{B|A})$ $ = \varepsilon$, then there exists $a \in \mathcal A$ such that $\mathbb V (P_{B|A=a},P'_{B |A= a} ) \leq 2 \varepsilon$, see~\cite[Lemma 4]{yassaee2014achievability}.
\end{enumerate}
\end{lem}

\vspace{1mm}
\begin{lem}[Coloring lemma $\mbox{\cite[Lemma 1]{csiszar1996almost}}$]\label{lem1csi}
Given a pair of random variables $(A, B)\in \mathcal A \times \mathcal B$ with joint distribution $P_{AB}$, marginals $P_A$ and $P_B$, and $\lvert \mathcal A \rvert \geq 4$, we have that
\begin{align*}
\frac{ { \mathbb V(P_{AB},P_{A}P_{B}) }^2}{2 \log2} &\leq I(A;B) \\[-3mm]
&\leq \mathbb V(P_{AB},P_{A}P_{B}) \log {\frac{\lvert \mathcal A\rvert }{\mathbb V(P_{AB},P_{A}P_{B})}},
\end{align*}
where the left-hand side is Pinsker's inequality.
\end{lem}

\vspace{1mm}
\begin{lem}[$\mbox{\cite[Lemma 5]{Cervia2017}}$]\label{lemmit}
Let $\bar P_{A}^{\otimes n}$ denote the i.i.d. product distribution associated with $\bar P_{A}$, and let $P_{A^{n}}$ be such that $\mathbb V(P_{A^{n}}, \bar P_{A}^{\otimes n}) \leq \varepsilon$.
Then, we have that 
\begin{equation*}
\sum\nolimits_{t=1}^{n} I(A_t;A_{\sim t}) \leq n\, f(\varepsilon),
\end{equation*}
for some function $f(\epsilon)$ where $\lim_{x\to 0}f(x)=0$.
In particular, if $P_{AB}$ is such that $\mathbb V(P_{AB}, \bar P_{A}\bar P_{B}) \leq \varepsilon$, 
then $ I(A;B) \leq f(\varepsilon) $.
\end{lem}


\section{System Model and Main Results}\label{sec: model}

\subsection{System Model}
\label{ssec:syst_model}


Consider the setting of Fig.~\ref{fig:wtap_model}, where a user observes an i.i.d. source $U^n$, with distribution $\bar P_{U}$, and wishes to \emph{coordinate} the actions of another user with $U^n$ by communicating through a discrete memoryless channel $\bar P_{YZ |X}$.
The communication is overheard by an eavesdropper, and thus the coordination must be achieved without leaking any information.

In this setting, coordination implies that the pair of sequences $(U^n,V^n)$ appears to be randomly drawn from the product distribution $\bar{P}_{UV}^{\otimes n}$.
The two legitimate users, an encoder and a decoder, share a source of common randomness $C \in \llbracket 1,2^{nR_0} \rrbracket$, which
is independent of the source $U^n$, to facilitate said coordination. 
The common randomness, $C$, may be thought of as being a secret key, shared between the legitimate users but unknown to the eavesdropper.
Therefore, it helps the encoder and the decoder coordinate their actions at the same time that allows secrecy.
As with a secret key, the legitimate users can obtain the common randomness from shared resources~\cite{bassi_2019_secretkey}.

The coordination scheme is as follows.
First, the encoder selects a signal $X^{n}= f_n(U^{n}, C)$, $f_n: \mathcal U^n \times \llbracket 1,2^{nR_0} \rrbracket \rightarrow \mathcal X^n$, which is transmitted over the channel $\bar P_{YZ|X}$ symbol by symbol.
Then, using the channel observation $Y^{n}$ and the common randomness $C$, the decoder selects an action $V^{n} = g_n(Y^{n}, C)$, $g_n: \mathcal Y^n \times \llbracket 1,2^{nR_0} \rrbracket \rightarrow \mathcal V^n$.
On the other hand, the eavesdropper observes the signal $Z^n$ but has no access to the common randomness $C$.
For block length $n$, the pair $(f_n , g_n )$ constitutes a \emph{coordination code}.

We want to understand the interplay between the strong coordination of the actions and secrecy: the induced joint distribution $P_{U^n V^n Z^n}$ has to satisfy the strong secrecy condition~\cite{elgamal2011nit} while strongly coordinating $(U^n, V^n)$.
We define the \emph{secure strong coordination} region as in~\cite[Definition 4.17]{cervia2018thesis}.

\vspace{1mm}
\begin{defi}[Secure strong coordination]
For a source $\bar P_{U}$ and a channel $\bar P_{YZ|X}$,
a pair $(\bar{P}_{V|U}, R_0)$ is achievable for secure strong coordination if there exists a sequence $(f_n,g_n)$ of encoders-decoders with rate of common randomness $R_0$, such that
\begin{align}
&\!\! \!\!\! \!  \lim_{n \to \infty} \mathbb V \left( P_{U^{n} V^{n}}, \bar{P}_{UV}^{\otimes n} \right)=0, \label{coorduvz}\\
&\!\! \!\!\! \!  \lim_{n \to \infty} \mathbb D(P_{U^nV^nZ^n} \Arrowvert P_{U^nV^n} P_{Z^n}) \!=\!\! \lim_{n \to \infty} I(U^nV^n;Z^n)\!=\! 0, \label{secrecyz}
\end{align}
where $P_{U^{n} X^{n} Y^{n} Z^n V^{n}}$ is the joint distribution induced by the source, the channel, and the coordination code.
The \emph{secure strong coordination region} $\mathcal{S}$ is the closure of the set of all achievable pairs $(\bar P_{V|U}, R_0)$.
\end{defi}

\vspace{2mm}

\begin{rem}
In the definition above, the first condition corresponds of \emph{strong coordination} of $U$ and $V$, and the second one is the \emph{strong secrecy} condition.
\end{rem}

\subsection{Main Results}

\subsubsection{General Case}
\label{sec:gen_case}

For the setting of Fig.~\ref{fig:wtap_model}, the secure strong coordination region is still unknown.
However, we provide here an inner bound for the region $\mathcal S$.

\vspace{1mm}
\begin{theo} \label{theoinout}
Let $\bar P_{U}$ and $\bar P_{YZ|X}$ be the given source and channel parameters, then $\mathcal S_{\text{in}} \subseteq \mathcal S$, where
\begin{align}
\mathcal S_{\text{in}} &\coloneqq \bigcup_{\bar{P}_{W_2 X}}
\begin{Bmatrix*}[l]
 (\bar P_{V|U}, R_0) : \\
 \quad \exists\, W_1\!\in\mathcal{W}_1, W_1\sim\bar P_{W_1|UV} \textnormal{ s.t. }\\
 \quad \bar P_{UW_1 V}=\bar P_{U} \bar P_{W_1|U} \bar P_{V| W_1} \\
 \quad I(W_1;U) \leq I(W_2;Y)\\
 \quad R_0 \geq I(UV;W_1) \\
 \quad \qquad\ - \big[ I(W_2;Y) - I(W_2;Z) \big] \\
 \quad \lvert \mathcal W_1 \rvert\leq \lvert \mathcal U \times \mathcal V \rvert +1
\end{Bmatrix*}, \label{eq:s_in}
\end{align}
and the union is with respect to all auxiliary random variables $W_2\!\in\mathcal{W}_2$ such that $\lvert \mathcal W_2 \rvert\leq \lvert \mathcal X \rvert +1$.
\end{theo}
\vspace{1mm}
\begin{IEEEproof}
The proof is deferred to Section~\ref{section ach gen}.
\end{IEEEproof}
\vspace{1mm}

In general, the region $\mathcal S_{\text{in}}$ is not tight due to the constraint related to the transmission of the auxiliary random variable $W_1$ through the channel, i.e., $I(W_1;U) \leq I(W_2;Y)$.
A similar issue is also encountered in~\cite{bassi_2016_secretkey}, albeit for the problem of secret key generation.
Nonetheless, we are able to characterize the region $\mathcal{S}$ in some cases, as we see next.

\vspace{1mm}
\begin{rem}
Observe that we use two random variables in this proof. Similarly to~\cite{Cervia2018journal}, this is due to the double purpose of the problem as in: coordinate the actions $U$ and $V$, and make their joint distribution independent from the observation of an eavesdropper.
\end{rem}

\vspace{1mm}
\subsubsection{More Capable Channel and Reliable Source Reconstruction}\label{sec:special_case}

Suppose that the decoder wants to reconstruct the source reliably, i.e., $\mathbb P \{U^n \neq V^n\} \leq f(\varepsilon)$, and  that the legitimate decoder's channel is more capable than the eavesdropper's, i.e., $I(X;Y) \geq I(X;Z)$ for all $\bar P_X$; then, the following result characterizes the secure strong coordination region.

\vspace{1mm}
\begin{cor}\label{deg theo}
If the decoder has a \emph{more capable channel} than the eavesdropper and \emph{reconstructs the source reliably}, the \emph{secure strong coordination region} is 
\begin{equation}\label{coord+degr}
\mathcal S \coloneqq \bigcup_{\bar P_X}
\begin{Bmatrix}[l]
(\bar P_{V|U}, R_0) :\\
 \quad \exists\, W\!\in\mathcal{W}, W\sim\bar P_{W|UV} \textnormal{ s.t. }\\
 \quad \bar P_{UWV}=\bar P_{U} \bar P_{W|U} \bar P_{V|W} \\
 \quad I(W;U) \leq I(X;Y)\\
 \quad R_0 \geq I(UV;W) \\
 \quad \qquad\ - \big[ I(X;Y) - I(X;Z) \big] \\
 \quad \lvert \mathcal W \rvert\leq \lvert \mathcal U \times \mathcal V \rvert +1
 \end{Bmatrix}.
 \end{equation}
\end{cor}
\begin{IEEEproof}
The proof is deferred to Section~\ref{sec:proof_specialcase}.
\end{IEEEproof}
\vspace{1mm}

\begin{rem}
We note that in the more capable case, as here, $C_{\textnormal{S}}\coloneqq$ $\max_{\bar P_X} \big[ I(X;Y)- I(X;Z) \big]$ is the secrecy capacity of the discrete memoryless wiretap channel~\cite[Thm.~22.1]{elgamal2011nit}. If we further assume all the channels to be symmetric, the uniform distribution 
$\bar P_X(x)=1/\lvert \mathcal{X} \rvert$, $\forall x \in \mathcal{X}$, maximizes the mutual information $I(X;Y)$ as well as the difference $I(X;Y)- I(X;Z)$. 
\end{rem}
\vspace{1mm}

The previous remark shows that, in our setting, the rate of common randomness is reduced by the presence of a secure channel. Instead of relying only on the common randomness $C$ for coordination, the encoder may generate random bits locally and transmit them securely to the decoder over the wiretap channel.

\subsection{Comparison with Previous Results}
\label{sec: compare}

In~\cite{cervia2018thesis}, the authors consider a two-node network, comprised of an information source and a noisy channel, and study the \emph{secure strong coordination} problem of an eavesdropper that receives the same sequence $Y^n$ as the decoder but has no knowledge of the common randomness $C$. 
Note that the capacity result in \cite[Thm.~4.15]{cervia2018thesis} presents the same rate constraints of Corollary~\ref{deg theo} if we specialize the latter to the case $Y=Z$. However, since in the assumption of Corollary~\ref{deg theo} we ask for reliable reconstruction of the source, the set of achievable target distributions is different.

For noiseless channels, a similar problem has been considered in~\cite{satpathy2016secure}, which analyzes a three-node cascade network, comprised of an i.i.d. source, two noiseless links, and a source of common randomness.
Note that the problem of finding the strong coordination region for the cascade setting is still open, but under the secrecy constraint that the source and the nodes' actions are independent of the messages exchanged via the noiseless links, 
\cite[Thm.~1]{satpathy2016secure} characterizes the secure strong coordination region. 
Similarly to \cite[Thm.~4.15]{cervia2018thesis}, we observe that by modifying Corollary~\ref{deg theo} with $Z=Y$, $W=U$, $V=Y$, $U=X$, and $I(X;Y)=R_1$, we recover the same rate conditions as in the special case of~\cite[Thm.~1]{satpathy2016secure} of
a cascade without the third node.

Observe that, in both~\cite{cervia2018thesis} and~\cite{satpathy2016secure}, a capacity region is derived when the eavesdropper observes the same sequence as the legitimate receiver, but without access to the common randomness.
With respect to the general case, \cite[Prop.~4.18]{cervia2018thesis} is a first attempt to provide an inner bound for the secure strong coordination region. In this paper, we present a more general achievability, as well as a converse for a special case.

\section{Proofs}\label{sec: gen proof}

In this part, we prove the inner bound for the secure strong coordination region in the general setting of Section~\ref{sec:gen_case} and the secure strong coordination region for the special setting of Section~\ref{sec:special_case}.

\subsection{Proof of Theorem~\ref{theoinout}}\label{section ach gen}

Let $\bar P_{W_2 X}$ be a distribution on $\mathcal{W}_2 \times \mathcal{X}$, and let $\mathcal S_{\text{in}}(\bar P_{W_2 X})$ be the region defined by the expression in curly braces on the r.h.s. of~\eqref{eq:s_in}.
In the sequel, we prove that for every $\bar P_{W_2 X}$ the region $\mathcal S_{\text{in}}(\bar P_{W_2 X})$ is achievable.

\vspace{1.5mm}
\subsubsection{Strong coordination of $(U^{n}, W^n, X^{n}, Y^{n}, Z^n, V^{n})$}\label{subsec: coord the seq}
 
We introduce an auxiliary random variable $W$ taking values in $\mathcal W$ such that 
\begin{equation}
\bar P_{UWXYZV}=\bar P_{U} \bar P_{W|U} \bar P_{X|UW} \bar P_{YZ|X} \bar P_{V|WY},
\label{eq:target_iid_pdf}
\end{equation}
where $\bar P_{X|UW}$ is a target i.i.d. conditional distribution.
The connection between $\bar P_{W|U} \bar P_{X|UW}$ in~\eqref{eq:target_iid_pdf} and $\bar P_{W_1| U} \bar P_{W_2 X}$ in the inner bound~\eqref{eq:s_in} will be made clear at the end of the proof.

Suppose that in the setting of Fig.~\ref{fig:wtap_model}, both the encoder and the decoder have access not only to the common randomness $C$ but also to an extra randomness $F$, where $C$ is generated uniformly at random in $\llbracket 1,2^{nR_0} \rrbracket$ with distribution $Q_C$ and $F$ is generated uniformly at random in $\llbracket 1,2^{n R} \rrbracket$ with distribution $Q_F$ and independently of $C$. 
Note that if we consider the marginal $\bar P_{UWXYV}$ of \eqref{eq:target_iid_pdf} with respect to $(U^{n}, W^{n}, X^{n}, Y^{n}, V^{n})$, \cite[Thm.~1]{Cervia2017} proposes an achievable strong coordination scheme.
The key idea is to consider two uniform (and independent) random binnings for $W^{n}$:
\begin{enumerate}\setlength{\itemsep}{0.2em}
\item binning $C = \varphi_1(W^{n})$, where $\varphi_1: \mathcal{W}^{n} \to \llbracket 1,2^{nR_0} \rrbracket$
maps each sequence of $\mathcal{W}^{n}$ uniformly and independently to the set $\llbracket 1,2^{nR_0} \rrbracket$;
\item binning $F = \varphi_2(W^{n})$, where $\varphi_2: \mathcal{W}^{n} \to \llbracket 1,2^{n R} \rrbracket$
maps each sequence of $\mathcal{W}^{n}$ uniformly and independently to the set $\llbracket 1,2^{nR} \rrbracket$.
\end{enumerate}
Then, a random binning and random coding schemes are presented, each of which induces a joint distribution, which we denote $P^{\text{RB}}$ and $P^{\text{RC}}$, respectively:
\begin{subequations}\label{eq:p_rb_rc}
\begin{align}
P^{\text{RB}} \coloneqq\ & \bar P_{U^{n}} \bar P_{W^{n}|U^{n}} \bar P_{X^{n}|W^{n} U^{n}} \bar P_{C|W^{n}} \bar P_{F|W^{n}} \nonumber\\
&\bar P_{Y^{n}|X^{n} } \bar P_{V^{n}|W^{n} Y^{n}} P^{\text{SW}}_{\hat W^{n}|C F Y^{n} },\\
P^{\text{RC}} \coloneqq\ &
Q_C Q_F \bar P_{U^{n}} \bar P_{ W^{n}|CFU^{n}} \bar P_{X^{n}|W^{n} U^{n}} \nonumber\\
&\bar P_{Y^{n}|X^{n}} P^{\text{SW}}_{\hat W^{n}|CF Y^{n}} \bar P_{V^{n}|\hat W^{n} Y^{n}}, 
\end{align}
\end{subequations}
where we added $\hat W^n$, i.e., the output of the Slepian--Wolf decoder $P^{\text{SW}}_{\hat W^{n}|CF Y^{n}}$. 

It is shown in~\cite{Cervia2017} that the distribution $ P^{\text{RB}}$ is trivially close in total variation distance to the target distribution $\bar P_{UWXYV}^{\otimes n}$. Moreover, because of the properties of random binning, if $ H(W|Y) < R+R_0 < H(W|U)$, the random binning and the random coding schemes have the same statistics, and thus $ P^{\text{RC}}$ is close in total variation distance to the target $\bar P_{UWXYV}^{\otimes n}$.

Now we observe that coordinating $Z^n$ as well as $(U^{n}, W^{n}, \allowbreak X^{n}, Y^{n}, V^{n})$ does not require more common randomness:
\begin{align}
\MoveEqLeft[1]
 \mathbb V \big(P^{\text{RB}}_{U^{n} X^{n} W^{n} \hat W^{n} Y^{n} Z^{n} V^{n} C F}, P^{\text{RC}}_{U^{n} X^{n} W^{n} \hat W^{n} Y^{n} Z^{n} V^{n} C F}\, \big) \nonumber\\
 &\quad \overset{\mathclap{(a)}}{=} 
 \mathbb V \big(P^{\text{RB}} P_{Z^{n}|X^{n} Y^{n}}, P^{\text{RC}} P_{Z^{n}|X^{n} Y^{n}} \big) \nonumber\\
 &\quad \overset{\mathclap{(b)}}{=} 
 \mathbb V \big(P^{\text{RB}}, P^{\text{RC}}\, \big) \nonumber\\[1mm]
 &\quad =\delta(n), \label{tv}
\end{align}
where $(a)$ follows from the chain rule and the definitions of $P^{\text{RB}}$ and $P^{\text{RC}}$ in~\eqref{eq:p_rb_rc}; and, $(b)$ is due to~\eqref{cuff17} in Lemma~\ref{tv prop}.

\vspace{1.5mm}
\subsubsection{Reducing the rate of common randomness---Strong coordination of $(U^n, Z^n, V^n)$}\label{subsec: reduce} 

We note that even though the extra common randomness $F$ is required to coordinate $(U^{n}, W^{n}, \allowbreak X^{n}, Y^{n}, Z^n, V^{n})$, we do not need it in order to coordinate only $(U^{n}, Z^{n}, V^{n})$.
Observe that by~\eqref{cuff16} in Lemma~\ref{tv prop}, equation~\eqref{tv} implies that 
\begin{align}
\mathbb V (P^{\text{RB}}_{U^{n} X^{n} Y^{n} Z^n V^{n} F}, P^{\text{RC}}_{U^{n} X^{n} Y^{n} Z^n V^{n} F}) =\delta(n), \label{convf2 wt 3} 
\end{align}
and thus
\begin{equation}\label{convf2 wt} 
\mathbb V (P^{\text{RB}}_{U^{n} Z^{n} V^{n} F}, P^{\text{RC}}_{U^{n} Z^{n} V^{n} F}) =\delta(n).
\end{equation}
Similarly to~\cite{yassaee2014achievability}, we reduce the rate of common randomness by having the two nodes agree on $F=f$, using \emph{randomness extraction} techiniques (see for instance~\cite[Chapter 17]{csiszar2011information}). 
To do so, we recall the following result, inspired by the discussion in~\cite[Section III.A]{pierrot2013joint}, and proved in this slightly different formulation in~\cite[Lemma 2.20]{cervia2018thesis}.

\vspace{1mm}
\begin{lem}[Channel randomness extraction for discrete memoryless sources and channels]\label{1.4.2}
Let $A^n$, distributed according to $P_{A^n}$, be a DMS and $P_{B^n|A^n}$, a DMC. 
Moreover, let $\varphi_n: \mathcal B^n \to \llbracket 1,2^{nR} \rrbracket$ be a uniform random binning of $B^n$, and $K\coloneqq \varphi_n(B^{n})$.
Then, if $R \leq H(B|A)$, there exists a constant $\alpha > 0$ such that
\begin{equation}\label{eq1lem1.4.2}
\mathbb E_{\varphi_n} \big[ \mathbb D (P_{A^nK} \Arrowvert P_{A^n} Q_K ) \big] \leq 2^{-\alpha n},
\end{equation}
where $Q_K$ is the uniform distribution on $\llbracket 1,2^{nR} \rrbracket$.
\end{lem}
\vspace{1mm}

We now apply Lemma~\ref{1.4.2} to the variables $B^{n}=W^{n}$, $K=F$, and $A^{n}= (U^{n}, Z^{n}, V^{n})$.
Thus, if $R < H(W| UZV)$, there exists a fixed binning (see~\cite[Lemma~2.2]{bloch2011physical}) such that
\begin{align}
\mathbb V (P^{\text{RB}}_{U^{n} Z^{n} V^{n} F}, Q_F P^{\text{RB}}_{U^{n} Z^{n} V^{n}})=\delta(n). \label{bin3 wt}
\end{align}
Combining~\eqref{convf2 wt} and~\eqref{bin3 wt}, with the help of the triangle inequality, we have that
\begin{align}
\mathbb V ( Q_F P^{\text{RB}}_{U^n Z^n V^n}, P^{\text{RC}}_{U^n Z^n V^n F } ) = \delta(n).\label{convf3 wt} 
\end{align}
Note that $ P^{\text{RC}}_{U^n Z^n V^n F}= Q_F P^{\text{RC}}_{U^n Z^n V^n|F}$ according to the definition of $P^{\text{RC}}$.
%
Then, by~\eqref{lem4} in Lemma~\ref{tv prop}, \eqref{convf3 wt} implies that there exists $f \in \llbracket 1,2^{n R} \rrbracket$ such that
\begin{equation}
\mathbb V (P^{\text{RB}}_{U^{n} Z^{n} V^{n}| F=f}, P^{\text{RC}}_{U^{n} Z^{n} V^{n}| F=f}) =\delta(n). \label{convfixf wt} 
\end{equation}
Therefore, by fixing $F=f$ and using common randomness $C$, we have strong coordination for $(U^n, Z^n, V^n)$.

\vspace{1.5mm}
\subsubsection{Rate constraints}

The preceding achievable scheme has imposed the following rate constraints:
\begin{align*}
& H(W|Y) < R+R_0 < H(W|U), \\
& R < H(W|UZV).
\end{align*}
Therefore, we obtain:
\begin{align*}
 & R_0 > H(W|Y)-H(W|UZV),\\
 & I(W;U) < I(W;Y). 
\end{align*}

\vspace{.5mm}
\subsubsection{Achieving secrecy---Independence between source and channel variables}

We now consider the case where the random variables of the channel are independent from the random variables of the source, i.e., $W=(W_1, W_2)$ and $(U, W_1, V)$ independent of $(W_2, X, Y, Z)$. Since
\begin{align*}
\MoveEqLeft[2]
  H(W|Y)-H(W|UZV) \\
  &= H(W_1 W_2|Y)-H(W_1 W_2|UZV)\\
  &= I(W_1 W_2; UZV) - I(W_1 W_2; Y)\\
  &= I(W_1 ; UV) + I(W_2 ;Z)- I(W_2 ; Y),
\end{align*}
the target distribution and information constraints become: 
\begin{align*}
 & \bar P_{U} \bar P_{W_1|U} \bar P_{V|W_1} \bar P_{W_2} \bar P_{X|W_2} \bar P_{YZ|X},\\
 & I(W_1;U)< I(W_2;Y), \\
 & R_0 > I(W_1 ; UV) + I(W_2 ;Z)- I(W_2 ; Y).
\end{align*}

Observe that in this setting, marginalizing over the uncoordinated variables, the target distribution is of the form $\bar{P}_{UV}^{\otimes n}\bar{P}_{Z}^{\otimes n}$. 
Therefore achieving strong coordination of the sequences $(U^n, Z^n, V^n)$ means that $\mathbb V \big( \bar{P}_{UV}^{\otimes n} \bar{P}_{Z}^{\otimes n}\!, P^{\text{RC}}_{U^nV^nZ^n} \big)$ vanishes.
By the upper bound on the mutual information in Lemma \ref{lem1csi}, the strong secrecy condition \eqref{secrecyz} is verified 
since there exists a sequence of codes such that $\mathbb V \big( \bar{P}_{UV}^{\otimes n} \bar{P}_{Z}^{\otimes n}\!, P^{\text{RC}}_{U^nV^nZ^n} \big)$ goes to zero exponentially.
Moreover, strong coordination of $(U^{n}, Z^{n}, V^{n})$ implies~\eqref{coorduvz}.

To conclude the proof of achievability, we note that only a part of the auxiliary random variable $W$, i.e., $W_1$, is used to coordinate the actions. The other part, $W_2$, is used to (implicitly) create a wiretap code for the channel.
Hence, for every fixed $\bar P_{W_2 X}$ we have proved that the region $\mathcal S_{\text{in}}(\bar P_{W_2 X})$ is achievable for secure strong coordination.
\endIEEEproof

\vspace{1mm}
\begin{rem}[Exponential speed of convergence]
The proof of the convergence in total variation distance relies on Lemmas~\ref{tv prop} and~\ref{1.4.2}. By Lemma~\ref{1.4.2}, the Kullback--Leibler divergence goes to zero exponentially, and so does the total variation distance (see the r.h.s. of Lemma~\ref{lem1csi}). On the other hand, the properties of the total variation distance metric in Lemma~\ref{tv prop} do not modify the speed of convergence.
\end{rem}


\subsection{Proof of Corollary~\ref{deg theo}}\label{sec:proof_specialcase}

\subsubsection{Achievability}
The achievability follows from the general inner bound $\mathcal S_{\text{in}}$, with $W_2=X$ and $W_1=W$.

\vspace{1.5mm}
\subsubsection{Converse}

According to the system model described in Section~\ref{ssec:syst_model}, let $C$ be a source of common randomness accessible to both the encoder and the decoder, and uniformly distributed in $\llbracket 1,2^{nR_0} \rrbracket$, for a sufficiently large $n$.
Consider a code $(f_n,g_n)$ that induces a distribution $P_{U^{n} V^{n}}$ on the actions that is $\varepsilon$-close in total variation distance to the i.i.d. distribution $\bar{P}_{UV}^{\otimes n}$ and such that $I(U^n V^n; Z^n)$ $\leq$ $f(\varepsilon)$. 
Furthermore, let the random variable $T$ be uniformly distributed over the set $\llbracket 1,n\rrbracket$ and independent of the variables $(U^{n}, X^{n}, Y^{n}, Z^n, V^{n}, C)$.

First, to obtain the bound on the information constraint, observe that
{\allowdisplaybreaks
\begin{align}
0 &\overset{\mathclap{(a)}}{\leq} I(X^{n};Y^{n}) - I(Y^{n}; U^{n} C) \nonumber\\[1mm]
 &\leq I(X^{n};Y^{n}) - I(Y^{n}; U^{n}| C) \nonumber \\
 &\overset{\mathclap{(b)}}{\leq} n\,I(X;Y) -\sum\nolimits_{t=1}^{n} I(Y^{n}; U_t|U^{t-1} C) \nonumber \\
 &= n\,I(X;Y) -\sum\nolimits_t \big[ I(Y^{n} U^{t-1} C; U_t) - I(U^{t-1} C; U_t) \big] \nonumber\displaybreak[2]\\
 &\overset{\mathclap{(c)}}{=} n\,I(X;Y) -\sum\nolimits_t I(Y^{n} U^{t-1} C; U_t) \nonumber\displaybreak[2]\\
 &\leq n\,I(X;Y) -\sum\nolimits_t I(Y^{n} C; U_t) \nonumber\displaybreak[2]\\
 &\overset{\mathclap{(d)}}{=} n \big[ I(X;Y) -I(W_{T}; U_T|T) \big] \nonumber\displaybreak[2]\\[1mm]
 &= n \big[ I(X;Y) -I(W_{T} T; U_T) + I(T; U_T) \big] \nonumber\displaybreak[2]\\
 &\overset{\mathclap{(e)}}{=} n \big[ I(X;Y) - I(W; U) \big], \label{converse2}
\end{align}}%
where $(a)$ follows from the Markov chain $ (U^n,C)-X^n-Y^n$;
$(b)$ is due to the channel $\bar P_{Y|X}$ being memoryless;
$(c)$ and $(e)$ follow from the i.i.d. nature of the source $\bar P_{U}$ and the independence of the source from the common randomness; and,
$(d)$ and $(e)$ follow from identifying the auxiliary random variables $W_{t}\coloneqq(C,Y^n)$ for each $t \in \llbracket 1,n \rrbracket$ and introducing the time-sharing random variable $T$, and $W\coloneqq (W_{T}, T)=(C,Y^n,T)$.

For the source of common randomness, we have that%
{\allowdisplaybreaks%
\begin{align}
 nR_0 &= H(C) \nonumber\\[.5mm]
 &\geq H(C|Y^n)- H(C|Y^n U^nV^n) \nonumber\\[.5mm]
 &= I(U^{n} V^n;C|Y^n) \nonumber\\[.5mm]
 &= I(U^{n} V^n; C Y^{n}) - I(U^{n} V^n; Y^{n}) \nonumber\\[-1mm]
 &\overset{\mathclap{(f)}}{\geq} I(U^{n} V^n; C Y^{n}) - I(U^{n} V^n; Y^{n} )\nonumber\\ &\quad+ I(U^{n} V^n; Z^{n} ) - f(\varepsilon),
 \label{eq:conv_cr}
\end{align}}%
where $(f)$ comes from the secrecy condition.
Now, the first term on the r.h.s. of~\eqref{eq:conv_cr} may be bounded as follows%
{\allowdisplaybreaks
\begin{align}
\MoveEqLeft[1]
I(U^{n} V^n; C Y^{n}) \nonumber\\
 &=  \sum\nolimits_{t=1}^n I(U_t V_t; C Y^{n}| U^{t-1} V^{t-1}) \nonumber\\[.5mm]
 &= \sum\nolimits_t \big[ I(U_t V_t ; C Y^{n} U^{t-1} V^{t-1} ) - I(U_t V_t ; U^{t-1} V^{t-1} ) \big] \nonumber\\[-1mm]
 &\overset{\mathclap{(g)}}{\geq} \sum\nolimits_t I(U_t V_t ; C Y^{n} U^{t-1} V^{t-1} ) -nf(\varepsilon) \nonumber\\[.5mm]
 &\geq \sum\nolimits_t I(U_t V_t ; C Y^{n} ) -nf(\varepsilon) \nonumber\\
 &\overset{\mathclap{(h)}}{=} n\, I(U_T V_T ; W_{T} |T ) -n f(\varepsilon) \nonumber\\[1mm]
 &= n \big[ I(U_T V_T ; W_{T} T ) - I(U_T V_T ; T ) \big] -n f(\varepsilon) \nonumber\\[-1mm]
 &\overset{\mathclap{(i)}}{\geq} n\, I(U_T V_T ; W_{T} T ) -2n f(\varepsilon) \nonumber\\
 &\overset{\mathclap{(j)}}{=} n\, I(U V ; W ) -2nf(\varepsilon),
\label{eq:conv_cr2}  
\end{align}}%
where $(g)$ and $(i)$ are due to Lemma~\ref{lemmit} and~\cite[Lemma VI.3]{cuff2013distributed}, respectively, since the induced distribution $P_{U^{n} V^{n}}$ is close to i.i.d. by hypothesis;
$(h)$ stems from the definition of the auxiliary random variable $W_{t}$ for each $t \in \llbracket 1,n \rrbracket$ and the introduction of the time-sharing random variable $T$; and,
$(j)$ follows from defining $U\coloneqq U_T$, $V\coloneqq V_T$, and $W\coloneqq (W_{T}, T)=(C,Y^n,T)$.

For the remaining terms on the r.h.s. of~\eqref{eq:conv_cr}, consider
{\allowdisplaybreaks
\begin{align}
\MoveEqLeft[1]
I(U^{n} V^n; Y^{n}) - I(U^{n} V^n; Z^{n}) \nonumber\\
 &\leq I(U^{n}; Y^{n}) - I(U^{n}; Z^{n}) + I(V^n; Y^{n}| U^{n}) \nonumber\\[.5mm]
 &\leq I(U^{n}; Y^{n}) - I(U^{n}; Z^{n}) + H(V^n| U^{n}) \nonumber\\[-1mm]
 &\overset{\mathclap{(k)}}{\leq} I(U^{n}; Y^{n}) - I(U^{n}; Z^{n}) + nf(\varepsilon) \nonumber\\  
&= \sum_{t=1}^n \big[ I(U^{n}; Y_t | Y^{t-1}) - I(U^{n} ; Z_t| Z_{t+1}^{n}) \big] +nf(\varepsilon) \nonumber\\[-1mm]
&\overset{\mathclap{(l)}}{=} \sum\nolimits_t \big[ I(U^{n} Z_{t+1}^{n}; Y_t | Y^{t-1}) \nonumber\\
&\qquad\qquad - I(U^{n} Y^{t-1} ; Z_t| Z_{t+1}^{n}) \big] + nf(\varepsilon) \nonumber\\
&\overset{\mathclap{(m)}}{=} \sum\nolimits_t \big[ I(U^{n};Y_t | Y^{t-1} Z_{t+1}^{n}) \nonumber\\
&\qquad\qquad - I(U^{n} ; Z_t| Y^{t-1} Z_{t+1}^{n} ) \big] + nf(\varepsilon) \nonumber\\
&\overset{\mathclap{(n)}}{=} \sum\nolimits_t \big[ I(W_{2,t};Y_t | W_{3,t}) - I(W_{2,t}; Z_t| W_{3,t}) \big] + nf(\varepsilon) \nonumber\\
&\overset{\mathclap{(o)}}{=} n \big[ I(W_{2}; Y_T | W_3) - I(W_{2} ;Z_T| W_3) \big] +nf(\varepsilon) \nonumber\\[1mm]
&\leq n\max_{i} \big[ I(W_{2} ; Y_T | W_3\!=\!i ) - I(W_{2}; Z_T| W_3\!=\!i ) \big] \nonumber\\
&\quad+nf(\varepsilon) \nonumber\\[-1mm]
&\overset{\mathclap{(p)}}{=} n \big[ I(\tilde W_2;Y)- I(\tilde W_2;Z) \big] + nf(\varepsilon) \nonumber\\[-1mm]
&\overset{\mathclap{(q)}}{\leq} n \big[ I(X;Y)- I(X;Z) \big] + nf(\varepsilon), \label{eq:conv_cr3}
\end{align}}%
where $(k)$ is due to Fano's inequality~\cite{fano1961transmission} and the assumption of reliable source reconstruction;
$(l)$ and $(m)$ stem from the Csisz{\'a}r sum identity~\cite{elgamal2011nit};
$(n)$ and $(o)$ follow from identifying the auxiliary random variables
$W_{3,t}\coloneqq (Y^{t-1}, Z_{t+1}^n)$, $W_3\coloneqq (W_{3,T}, T)=(Y^{T-1}, Z_{T+1}^n,T)$, $W_{2,t}\coloneqq (U^n, W_{3,t})$, and $W_2\coloneqq (W_{2,T}, T)=(U^n, W_3)$; and,
$(p)$ follows from defining $Y\coloneqq Y_T$, $Z\coloneqq Z_T$, and $\tilde W_2\sim P_{W_2|W_3=i^\star}$ (where $i^\star$ is the value that maximizes the bound), and the Markov chain $\tilde{W}_2-X-(Y,Z)$; this last step is possible since $\tilde{W}_2$ does not appear in any other bound.
Finally, $(q)$ is due to the more capable assumption.
Combining~\eqref{eq:conv_cr}--\eqref{eq:conv_cr3}, we obtain the bound on the rate of the source of common randomness.

Similarly to~\cite{cuff2013distributed}, the cardinality bound on $\mathcal{W}$  is a consequence of the Fenchel--Eggleston--Carath\'eodory's theorem~\cite[Appendix C]{elgamal2011nit}; and the proof is omitted.
\hfill\IEEEQED

\bibliographystyle{IEEEtran}
\bibliography{IEEEabrv,mybib}

\end{document}